\documentclass[aps,amsmath,prb,twocolumn,showpacs]{revtex4-1}
\usepackage{graphics}
\usepackage{graphicx}
\usepackage{amsmath}
\usepackage{amsfonts}
\usepackage{bm}
\usepackage{color}
\usepackage{bbm}

\newcommand{\be}{\begin{equation}}
\newcommand{\ee}{\end{equation}}
\def\boldsymbol#1{\mbox{\boldmath$#1$}}
\newcommand{\p}{\mathbf{p}}

\newcommand{\beq}{\begin{equation}}
\newcommand{\eeq}{\end{equation}}
\newcommand{\beqa}{\begin{eqnarray}}
\newcommand{\eeqa}{\end{eqnarray}}
\newcommand{\bea}{\begin{eqnarray}}
\newcommand{\eea}{\end{eqnarray}}
\usepackage{amssymb}
\usepackage{array}
\usepackage{amsmath}
\usepackage{graphicx}\usepackage{graphics}
\usepackage{dcolumn}
\usepackage{bm}\usepackage{varioref}

\begin{document}

\title{Spiraling Fermi arcs in Weyl materials}

\author {Songci Li}

\affiliation{Department of Physics, University of Washington, Seattle, WA 98195}

\author{A. V. Andreev}

\affiliation{Department of Physics, University of Washington, Seattle, WA 98195}

\date{\today}

\begin{abstract}
In Weyl materials the valence and conduction electron bands touch at an even number of isolated points in the Brillouin zone. In the vicinity of these points the electron dispersion is linear and may be described by the massless Dirac equation. This results in nontrivial topology of Berry connection curvature. One of its consequences is the existence of peculiar surface electron states whose Fermi surfaces form arcs connecting projections of the Weyl points onto the surface plane. Band bending near the boundary of the crystal also produces surface states. We show that in Weyl materials band bending near the crystal surface gives rise to spiral structure of energy surfaces of arc states. The corresponding Fermi surface has the shape of a spiral that winds about the projection of the Weyl point onto the surface plane. The direction of the winding is determined by the helicity of the Weyl point and the sign of the band bending potential. For close valleys arc state morphology may be understood in terms of avoided crossing of oppositely winding spirals.
\end{abstract}


\maketitle

The possibility of existence of zero band gap semiconductors with linear electron dispersion near the contact points of conduction and valence bands in the Brillouin zone and the stability of this state with respect to Coulomb interactions has been investigated extensively beginning with the work of Abrikosov and Beneslavskii.~\cite{Beneslavskii} Recently there has been a renewal of theoretical~\cite{Qi,Vafek,Wehling,Murakami,Balents,KaneMele,
Hasan,Ashvin,Burkov,Aji,Spivak,Okugawa,Y.Sun,Haldane} and experimental research~\cite{LingLu,Ong,GenfuChen,Hasan1,Hasan2,B. Q. Lv,Hasan3,L. X. Yang} on Weyl meterials, in which the electron dispersion near the point of crossing of the two non-spin-degenrate bands corresponds to that of relativistic  massless (Weyl) fermions. The corresponding Berry connection curvature~\cite{Niu} is divergence free with the exception of the band contact points, where monopole ``charges''  are located. By  Nielsen and Ninomiya~\cite{Ninomiya} such Weyl point come in pairs of opposite helicity. In Weyl materials the  Adler-Bell-Jackiw chiral anomaly,~\cite{Adler,BellJackiw} first discovered in particle physics, may be realized. This gives rise to giant anisotropic negative magnetoresistance.~\cite{Ninomiya,Spivak,Ong,GenfuChen}

Weyl materials possess surface states~\cite{Vafek,Ashvin}
whose Fermi surfaces have the shapes of arcs connecting the projections of the Weyl points onto the crystal plane of the surface. Their origin is also rooted in the nontrivial topology of the Berry phase connection.~\cite{Haldane} Arc states are close relatives of chiral edge states in graphene.~\cite{Nakada,Fujita,Hatsugai} The latter also owe their existence to the structure of Berry curvature in graphene,~\cite{Hatsugai}, and their spectrum (as a function of momentum along the edge) consists of a line connecting the projections of the two Dirac points onto the edge. Evidence for arc states was recently observed by angle resolved photoemission spectroscopy (ARPES) in several Weyl materials.~\cite{Hasan1,Hasan2,B. Q. Lv,Hasan3,L. X. Yang}

\begin{figure}
	\includegraphics[height=2.0in,width=3.2in]{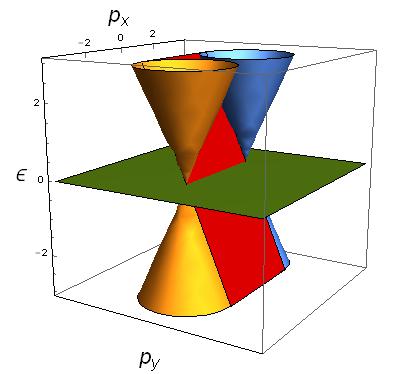}
	\caption{\label{fig:diraccone&sprial} Sketch of the energy dispersion of the bulk states (interior of the Dirac cones) and the surface arc states (inclined plane segment) in the absence of band bending.}	
\end{figure}

Another common mechanism of surface state generation is band bending. It is caused by the difference between electron environments at the surface and in the bulk. In previous theoretical considerations of arc states breaking of particle-hole symmetry due to band bending near the surface was ignored (breaking of particle-hole symmetry at the graphene boundary leads to a finite dispersion of these, otherwise dispersionless, surface states.~\cite{Akhmerov1,Chen,Akhmerov2}) The resulting energy dispersion of arc states is depicted in Fig.~\ref{fig:diraccone&sprial}.

\begin{figure}[t]
	\begin{center}
		\includegraphics[height=1.8in,width=3.5in]{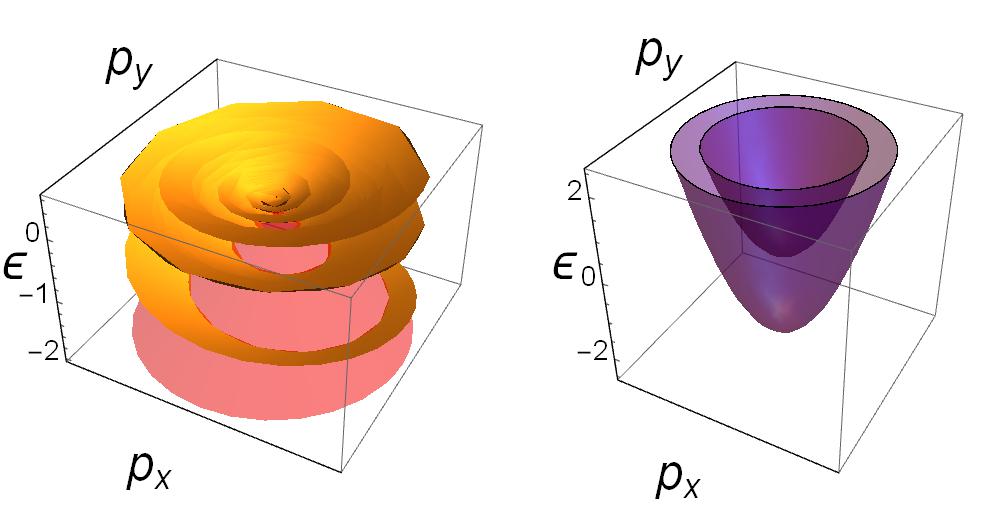}
		\caption{\label{fig:surfacecoulomb} Left: the spiral energy surface of bound states for band bending potentials obtained from Eq.~(\ref{eq:U_solution}) with parameters $\gamma=4.5,\ z_0=0.01$. For clarity only the part with $\theta \in (-\pi/2, \pi/10)$ is shown. The surface terminates at the Dirac cones of particle-like and hole-like states of the continuum.  Right: parabolic energy surfaces of bound states for a conventional semiconductor.}
	\end{center}
\end{figure}

\begin{figure}
	\includegraphics[height=1.7in,width=3.0in]{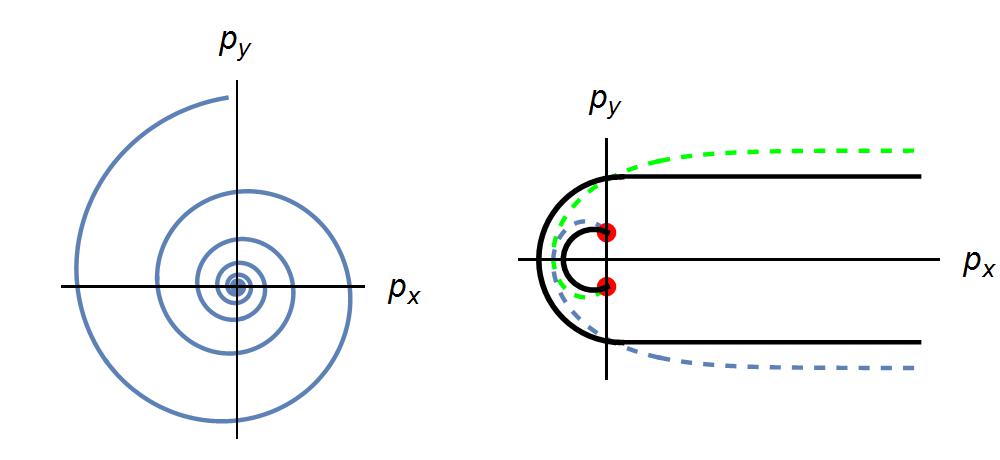}
	\caption{\label{fig:spiral} Left: Fermi arc shape for the undoped Weyl semiconductor determined from Eq.~(\ref{eq:boundstate}). The spiral makes infinite number of turns about the Weyl point. Right: Dashed lines -- Fermi arc shapes for two nearby valleys of opposite helicity for a square well potential, Eq.~(\ref{eq:squarewell}). The spirals make a finite number of turns (equal to the number of the bound states) about the Weyl points (red dots). Solid lines -- avoided crossing of the two spirals due to finite valley mixing.}	
\end{figure}

In this letter we study the effects of band-bending near the surface of Weyl materials. For conventional semiconductors surface states caused by band bending are characterized by closed Fermi surfaces, see right panel in Fig.~\ref{fig:surfacecoulomb}. We show that in Weyl materials surface states caused by band bending hybridize with the arc states to form a single band with spiraling energy dispersion, shown in the left panel of Fig.~\ref{fig:surfacecoulomb}. The corresponding Fermi surface is shaped as a spiral that winds about the projection of the Weyl points onto the boundary, see the left panel of Fig.~\ref{fig:spiral}.
%
%

\begin{figure}
	\includegraphics[height=1.5in,width=2.5in]{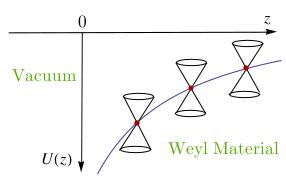}
	\caption{\label{fig:bandbending} Schematics of band bending potential $U(z)$ and Dirac cones close to the boundary.}	
\end{figure}

We consider a crystal boundary perpendicular to the $z$-axis, as illustrated in Fig.~\ref{fig:bandbending}.
In the two band approximation an electron state with a fixed crystalline momentum parallel to the boundary, $\p$,  may be described by a two-component pseudospinor,
\[
\psi_\p(z)=\left(
       \begin{array}{c}
         u_\p(z) \\
         v_\p(z) \\
       \end{array}
     \right).
\]
Assuming for simplicity of presentation an isotropic~\cite{Anisotropy} dispersion near the Weyl points we write the effective low energy Hamiltonian in the form
\begin{equation}\label{eq:h_Dirac}
  \hat H=U(z) + v \p \cdot \boldsymbol{\sigma} - i \hbar v \partial_z \sigma_z.
\end{equation}
Here $\p=(p_x,p_y)$ and $\boldsymbol{\sigma}$ are Pauli matrices, and the band bending  is described by the potential $U(z)$.
The eigenstates with energy $\epsilon$  obey the corresponding Dirac equation,
\begin{equation}\label{eq:Dirac_article}
  \left(
    \begin{array}{cc}
      \epsilon+ i \hbar v \partial_z - U(z) & -v p \\
       - v\bar p & \epsilon- i \hbar v \partial_z - U(z) \\
    \end{array}
  \right)\left(
           \begin{array}{c}
             u_\p(z) \\
             v_\p(z) \\
           \end{array}
         \right)=0,
\end{equation}
where $p=p_x-i p_y$,  and $\bar p=p_x+ i p_y$.

In a general situation the projections of different Weyl points onto the plane of momentum parallel to the boundary do not coincide. In this case the boundary conditions do not couple  different valleys and can be parameterized by a single phase $\chi$,
\begin{equation}\label{eq:bc_chi}
  \left(v_\p-e^{i\chi}u_\p\right)\Big|_{z=z_0}=0.
\end{equation}
This corresponds to a vanishing current normal to the boundary, $j_z= v \psi^\dagger \sigma_z \psi$.~\cite{Akhmerov1,Akhmerov2}

The pseudospinor amplitudes obey second order differential equations, i.e.
\begin{equation}\label{eq:u_eq_article}
  u''_\p (z)+\Bigg\{\left[\frac{\epsilon -U(z)}{\hbar v}\right]^2 + i \frac{U'(z)}{\hbar v} - \frac{|\p|^2}{\hbar^2}\Bigg\} u_\p(z)=0.
\end{equation}
Here the derivative with respect to $z$ is denoted by prime.
As follows from (\ref{eq:Dirac_article}) the $v$ amplitude may be obtained from the $u$ amplitude by a differential operation
\begin{equation}\label{eq:v_u}
  p v_\p(z)=\left[\epsilon - U(z)+ i \hbar v \partial_z\right]u_\p(z).
\end{equation}
The momentum dispersion of states bound to the surface, $\epsilon(\p)$, is determined by discrete spectrum of the operator in Eq.~(\ref{eq:h_Dirac}) with the boundary condition Eq.~(\ref{eq:bc_chi}). Although our conclusions regarding the spiral structure of the dispersion  of surface states, $\epsilon(\p)$, are quite general, the specific shape of the dispersion depends on the details of the band-bending potential.

The energy dispersion relation near the boundary of the continuum depends on the long (as compared to the lattice constant) distance behavior of the potential $U(z)$. The corresponding macroscopic electric fields are created by the double layer charge distribution near the surface and depend on the screening properties of the Weyl material. In this respect one needs to distinguish between two cases: i)
For zero band gap semiconductors at zero temperature, $T=0$, linear screening is absent and $U(z)$ is expected to decay as a power of $z$, ii) For zero band gap semiconductors at finite temperature or semimetals,~\cite{note1} $U(z)$ decays exponentially at distances greater that the screening length (which is much longer than the lattice constant).
In either of these cases the spatial dependence of $U(z)$ at long distance may be analyzed using the Poisson's equation, $U''(z)=-4\pi e^2 n(z)$, where $e$ is the electron charge and $n(z)$ is the density of electrons.

\textit{Semiconducting~\cite{note1} case.} In this case the thermodynamic electronic density of states (compressibility) $\partial n/\partial\mu$ vanishes in the bulk for an undoped system. As a result, linear screening is absent, and the potential $U(z)$ exhibits a power law fall off with the distance $z$ from the boundary.

For small coupling constants, $\alpha=e^2/\hbar v$, the band-bending region near the surface hosts many bound states. In this case the spatial dependence of the electron density may be determined using the Thomas-Fermi approximation. In the undoped crystal the electrochemical potential is measured with respect to the Weyl point, i.e. $U(z) +\mu(z)=0$. Due to the linear dispersion the electron density depends on the local potential $U(z)$ as $n(z)=-g U(z)^3/3\pi^2\hbar^3v^3$, where $g$ is the number (even) of the Weyl points in the Brillouin zone.  This yields the Thomas-Fermi equation in the form
\begin{equation}\label{eq:Thomas-Fermi}
  U''(z)=\frac{4g\alpha}{3\pi(\hbar v)^2}U^3(z).
\end{equation}
The relevant solution is given by
\begin{equation}\label{eq:U_solution}
   U(z)=\hbar v\frac{\gamma}{z}, \quad |\gamma|= \sqrt{\frac{3\pi}{2g\alpha}}\gg 1.
\end{equation}
The short-distance behavior of the confining potential may not be understood within the low energy framework employed here and is accounted for by the boundary condition (\ref{eq:bc_chi}), which should be imposed at distance of the order of the lattice constant.

The second order differential equation Eq.~(\ref{eq:u_eq_article}) with the potential $U(z)$ given by Eq.~(\ref{eq:U_solution}) can be reduced to the equation for the confluent hypergeometric function. Introducing the dimensionless variable $\zeta= 2 \sqrt{|\p|^2 -(\epsilon/v)^2}\, \, z/\hbar$ we can write the general solution of Eq.~(\ref{eq:u_eq_article}) in the form
\begin{equation}\label{eq:u_w_article}
  u(\zeta) = \zeta^{-i\gamma} \exp (- \zeta/2) w(\zeta),
\end{equation}
were $w(\zeta)$ is expressed in terms of  the confluent hypergeometric function $\Phi(a,c\,; \zeta)$~\cite{Erdelyi_vol_1} as
\begin{equation}\label{eq:w_solution_article}
w(\zeta)= \Phi\left(a, c\,; \zeta\right) + C_1\,  \zeta^{1-c} \Phi\left(a-c+1, 2-c\,;\zeta\right)
\end{equation}
with $
  a=\gamma\left(\epsilon/\sqrt{v^2|\p|^2 -\epsilon^2} - i\right) $, and $  c=-2i\gamma$.

For bound states the value of the constant $C_1$ is determined from the condition that the wavefunction decay exponentially at $\zeta \to \infty$. Using the large distance asymptotic of $\Phi(a, c\,;\zeta)$, we find
\begin{equation}\label{eq:C_1}
       C_1=\frac{a^*}{c(1-c)}
    \frac{\Gamma(a^*)}{\Gamma(a)}
    \frac{\Gamma(c)}{\Gamma(c^*)},
\end{equation}
where $\Gamma(x)$ is the gamma function.
The spectrum of surface states is determined by substituting this form of $u$ given by Eqs.~(\ref{eq:u_w_article})-(\ref{eq:C_1}) into Eq.~(\ref{eq:v_u}) and using the boundary condition, Eq.~(\ref{eq:bc_chi}). Since the latter is imposed at short distances $z=z_0\ll \hbar/|\p|$, which corresponds to $\zeta_0 \ll 1$,
the result of this substitution simplifies to
\begin{equation}\label{eq:boundstate}
e^{i \left(\phi-  \theta - \frac{\pi}{2}\right)}\frac{\Gamma(a^*)\Gamma(c)}{\Gamma(a)\Gamma(c^*)}\,  e^{2i\gamma \ln\left(\frac{ 2|\p|z_0}{\hbar}\cos\theta\right) }=
e^{i\chi}.
\end{equation}
Here $a=\gamma(\tan\theta-i)$, and we parameterized the energy of bound states, $|\epsilon| < v |\p|$, by the angle $\theta$, so that $
  \epsilon=v|\p|\sin\theta$, with $\theta\in(-\pi/2, \pi/2)$,
and introduced the momentum azimuthal angle $\phi$, $\p= |\p|(\cos \phi, \sin \phi)$.

Eq.~(\ref{eq:boundstate}) has a clear interpretation. Vanishing of the normal current $j_z=v\psi^\dagger \sigma_z \psi$ in a stationary state implies that the ratio $v_\p(z)/u_\p(z)$ is a pure phase factor. The condition of exponential decay of the solution at  $z\to \infty$ requires that $v_\p(z)/u_\p(z)|_{z\to \infty}=e^{i \left(\phi+ \theta - \frac{\pi}{2}\right)}$. The remaining factor in the left hand side of Eq.~(\ref{eq:boundstate}) describes the dynamical phase accumulated during the motion between the boundary and $z\to \infty$. By the boundary condition (\ref{eq:bc_chi}) the product of these phase factors must be equal to $e^{i\chi}$ in the right hand side.

In the absence of the band bending potential the dynamical phase vanishes. In this case bound states exist only in half the momentum plane, $\phi-\chi \in (0, \pi)$, in agreement with the treatments ignoring the band bending effects near the crystal surface. At $\gamma \neq 0$ the dynamical phase accumulated due to the presence of the band bending potential allows the phase difference $\phi-\chi$ to depart from the interval $(0, \pi)$ and produces the spiraling structure of the energy surface. By Levinson's theorem~\cite{Landau3} this phase is given by $2\pi$ times the number of bound states in the corresponding confining potential. As a result, the number of turns made by the spiraling energy surface around $|\p|=0$ is equal to the number of bound states caused by band bending.

Due to the slow $1/z$ fall off, the potential in Eq.~(\ref{eq:U_solution}) hosts an infinite number of shallow bound states. This is reflected in the logarithmic divergence of the phase on the left hand side of Eq.~(\ref{eq:boundstate}) at small parallel momenta, $|\p| \to 0$. The shape of the Fermi surface of bound states in an undoped Weyl material is determined by setting $\theta=0$ (zero energy) in Eq.~(\ref{eq:boundstate}). The resulting spiral, which makes an infinite number of turns in the $p_x-p_y$ plane, is plotted in the left panel of Fig.~\ref{fig:spiral}.

The spectrum of bound states with nonzero energy is given by Eq.~(\ref{eq:boundstate}) with $\theta \neq 0$.
For each momentum  $\p=|\p|(\cos \phi, \sin \phi)$ it yields a discrete spectrum of bound states.  This defines a surface in three dimensional space with coordinates $p_x$, $p_y$ and $\epsilon$. In order to visualize this surface we can view Eq.~(\ref{eq:boundstate}) as the expression for the azimuthal angle $\phi$ as a function of energy ($\theta$) and the absolute value of momentum $|\p|$. This gives a parametric representation of the dispersion relation of bound states. A three dimensional parametric plot of $p_x=|\p|\cos\phi$, $p_y=|\p|\sin\phi$ and $\epsilon=v|\p|\sin\theta$ as a function of two parameters, $|\p|>0$ and $\theta\in (-\pi/2,\pi/2) $ that follows from Eq.~(\ref{eq:boundstate}) is presented in the left panel of Fig.~\ref{fig:surfacecoulomb}.

\textit{Semimetallic~\cite{note1} case.}
For potentials $U(z)$ that decay faster than $1/z$ into the bulk the number of bound states is finite. Therefore, the spiral Fermi surface makes a finite number of turns about the $\p=0$ point. As an illustration we consider the simplest example of such a potential, a square well of depth $U_0$, $U(z)= - U_0\theta (z_0-z)$. The treatment of this case as outlined above is straightforward and leads to the following counterpart of Eq.~(\ref{eq:boundstate}),
\begin{equation}\label{eq:squarewell}
e^{i\left(\phi+\theta -\frac{\pi}{2}\right)} \, \frac{q+(|\p|\cos\theta/\hbar+i\tilde{U})\tan(qz_0)}{q+(|\p|\cos\theta/\hbar-i\tilde{U})\tan(qz_0)}= e^{i\chi},
\end{equation}
where $\tilde{U}=U_0/\hbar v$ is measured with respect to wave vector and $q=\sqrt{\tilde{U}^2+2\tilde{U}|\p|\sin\theta/\hbar-(|\p|\cos\theta/\hbar)^2}$ , with the same parameterization as before. The shape of the Fermi surface can be obtained by setting $\theta$ to 0 in Eq.~(\ref{eq:squarewell}) and is plotted in the right panel of Fig.~\ref{fig:spiral}. The number of windings it makes in the $p_x-p_y$ plane is set by the number of bound states in the well, $\propto \tilde{U} z_0$. The characteristic momentum scale $p_s$ of the spiraling region is set by the depth of the well, $p_s\approx U_0/v$.

Note that the direction of the winding of the spiraling energy surface is opposite for  the opposite helicity of the Weyl valley or opposite sign of the  band bending potential. Thus for a known valley helicity the winding direction of the spiraling arc may be used to infer the sign of the band bending potential.

\emph{Close valleys.} The Fermi arcs join Weyl valleys with opposite helicity. If the momentum scale of the helical structure, $p_s\approx U_0/v$ is smaller than the momentum distance $\Delta p$ between the (projections on the terminating surface)  Weyl points then the spiral structures near each valley bay be considered independently, as above. In the opposite case, $p_s \lesssim \Delta p$ the interaction between the valleys becomes important. In many of the Weyl materials Weyl valleys form nearby pairs with intervalley distance $\Delta p$ significantly smaller than the Brillouin zone size.  In this case the effect of the intervalley interaction on the helical structure of arc states may be studied using the Dirac equation. Valley mixing of surface states arises from: i) valley coupling in the bulk, and ii) intervalley scattering from the crystal surface. The first effect is captured
by the four band Dirac Hamiltonian, so that Eq.~(\ref{eq:h_Dirac}) may be replaced with~\cite{KaneMele,Murakami,Balents} $\hat{H}=U(z) + v \tau_3 \mathbf{p} \cdot \boldsymbol{\sigma} + \Delta \tau_1\sigma_x$ with $\boldsymbol{\tau}$ being the Pauli
matrices in the valley subspace. In this formulation the current operator becomes $\mathbf{j}=e v\Psi^\dagger \tau_s \boldsymbol{\sigma}\Psi$. The intervalley scattering at the boundary is accounted for by generalizing the boundary condition  (\ref{eq:bc_chi}) that corresponds to the vanishing of the current $j_z$ normal to the boundary,
\begin{equation}\label{eq:bc_two_valleys}
    \vec{v}_\mathbf{p}= \hat{M} \vec{u}_\mathbf{p}.
\end{equation}
Here  $\vec{u}_\mathbf{p}$ and $\vec{v}_\mathbf{p}$ are the two component (in the valley subspace) generalizations of the spinor amplitudes $u_\mathbf{p}$ and $v_\mathbf{p}$,
and $\hat{M}$ is a $2\times 2$ unitary matrix whose nondiagonal elements describe intervalley scattering from the boundary. In the absence thereof the matrix reduces to the diagonal form (\ref{eq:bc_chi}), $M_{ij}=\delta_{ij}e^{i\chi_j}$.

In the absence of valley mixing the arc states form two intersecting spirals winding in opposite directions (due to the opposite helicity of the Weyl valleys), shown in the solid lines in the right panel in Fig.~\ref{fig:spiral}. Arc state morphology at weak mixing may be understood in terms of avoided crossing of these spirals. It yields two alternative morphologies of arc states Fermi surfaces. Type I -- all Fermi lines confined to the vicinity of the pair of the Weyl nodes (with one arc connecting them and, possibly one or more closed lines nearby). Type II -- a pair of arcs emanating from the coupled Weyl nodes to other Weyl pairs. In the four band Dirac model both cases are realized depending on the choice of parameters in the Hamiltonian and the boundary condition (\ref{eq:bc_two_valleys}).

To illustrate the effects of valley mixing let us for simplicity neglect the valley coupling $\Delta$ in the bulk. In this case the ratio of the spinor amplitudes $v^j_\p/u^j_\p$ in valley $j$ is a pure phase factor.  For the potentials we considered, its value at the boundary, $e^{i \alpha^j_\p}$, is given by the left hand side in Eqs.~(\ref{eq:boundstate}) and (\ref{eq:squarewell})~\cite{note2}. Thus, according to the boundary condition (\ref{eq:bc_two_valleys}) the spectrum of arc states  is determined by the equation $\det \left[\delta_{ij}e^{i \alpha^j_\p}- M_{ij} \right] =0$. Parameterizing the matrix $\hat{M}$ in Eq.~(\ref{eq:bc_two_valleys}) as $\hat{M}=e^{i \chi} e^{i \tau_2 \Theta}$, where the angle $\Theta$ describes the valley mixing, and $\chi$ is the phase characterizing the intra-valley reflection, c.f. Eq.~(\ref{eq:bc_chi}), and using the phase factors $e^{i \alpha^j_\p}$ for the square well potential, Eq.~(\ref{eq:squarewell}) we get the energy dispersion of mixed valleys. At zero valley mixing, $\Theta=0$ the Fermi arcs are shown by the dashed lines in the right panel of Fig.~\ref{fig:spiral}. Avoided crossings due to small valley mixing  results in type II morphology (solid lines), which is consistent with ``tadpole'' arc shape observed in Refs.~\onlinecite{Hasan1,Hasan2,L. X. Yang,B. Q. Lv,Hasan3}.

In summary, we studied the interplay between the conventional, band bending mechanism of surface state formation, and the topological mechanism leading to formation of arc states. It results in a spiral structure of energy dispersion of surface states near the Weyl nodes. For close valleys, the arc state morphology may be understood in terms of avoided crossing of the spirals of individual valleys. Our findings significantly affect the density and the dynamics of electrons in the surface metallic channel.

We are grateful to D. Cobden and B. Spivak for useful discussions. This work is supported by the U.S. Department of Energy Office of Science, Basic Energy Sciences under award number DE-FG02-07ER46452.


\begin{thebibliography}{99}
\bibitem{Anisotropy} Accounting for the anisotropy of the Dirac cones does not change our conclusions.
\bibitem{note1} To avoid confusion we note that following the traditional nomenclature we use the term semimetals for materials in which the zero temperature densities of electrons and holes are finite and equal to each other. In a Weyl material this corresponds to energy offset of some Weyl valleys relative to others, causing partial band overlap. The semiconducting case refers to materials with no energy offset between different valleys. In the literature on Weyl materials, in contrast to the conventional terminology,  these two cases are often referred to as respectively Weyl metals and Weyl semimetals.
\bibitem{note2} Note that the azimuthal angle $\phi$ and $|\p|$ in each valley should be determined relative to the corresponding Weyl node.
\end{thebibliography}
\end{document}